\documentclass[pra,showpacs,floatfix,preprint,amsmath,amssymb]{revtex4}

\usepackage{epsfig}
\usepackage{CJK}
\usepackage{amsmath}
\usepackage{supertabular}
\usepackage{graphicx}% Include figure files
\usepackage{dcolumn}% Align table columns on decimal point
\usepackage{bm}% bold math
\usepackage{float}
\usepackage{subfigure}
\begin{document}

\title{Scaling of quantum correlation and monogamy relation near a quantum phase transitions in two-dimensional XY spin system}% Force line breaks with \\

\author{Meng Qin$^{1,2}$$\footnote{$^{1}$Department of Physics and Key Laboratory of Modern Acoustics, Nanjing University, Nanjing 210093, China. $^{2}$College of Sciences, PLA University of Science and Technology, Nanjing, 211101, China. $^{3}$Center of Theoretical Nuclear Physics, National Laboratory of Heavy-Ion Accelerator, Lanzhou 730000, China. Correspondence and requests for materials should be addressed to M.Q. (email: qrainm@gmail.com) or Z.-Z.R. (email: zren@nju.edu.cn)}$, Zhong-Zhou Ren$^{1,3}$ $\&$ Xin Zhang$^{1}$}

\affiliation{ }

\noindent

\begin{abstract}
The purpose of the paper is mainly to investigate the quantum critical behavior of two-dimensional XY spin system by calculating quantum correlation and monogamy relation through implementation of quantum renormalization group theory.
Numerical analysis indicates that quantum correlation as well as quantum nonlocality can be used to efficiently detect the quantum critical property in two-dimensional XY spin system. The nonanalytic behavior of the first derivative of quantum correlation approaches infinity and the critical point is reached as the size of the model increases. Furthermore, we discuss the quantum correlation distribution in this model based on square of concurrence (SC) and square of quantum discord (SQD). The monogamous properties of SC and SQD are obtained for the present system. We finally reveal that the monogamy score can be used to capture the quantum critical point.

\end{abstract}

\maketitle

In order to show the nonlocality in quantum mechanics, Einstein, Podolsky, and Rosen proposed the thought experiment known as the EPR paradox in 1935 \cite {einstein}. Such kind of nonlocality is defined as entanglement later. The investigation on the nonlocality of quantum physics form a new discipline namely quantum information science. It is generally recognized that entanglement is the key resource in the discipline \cite {chen}. In recent years, the study on the monogamy of entanglement attract much attention, i.e., quantum entanglement cannot be freely shared among the constituents of a multipartite system \cite {qin, coffman}. Monogamy is one of the basic rules in making quantum cryptography secure and plays an indispensable role in superdense coding \cite {kumar}. Many achievements are obtained after introduced by Coffman et al. However, according to recent progress, there are also some states containing quantum correlation beyond entanglement and are also very effective in quantum information processing \cite {knill}. So, entanglement can not signify all the quantum nonlocality in a quantum system. Quantum correlation may be the most fundamental resource in quantum information protocols. Therefore, monogamy of quantum correlation also gets much attention \cite {bai, prabhu,liu}. Motivated by the development of monogamy of quantum correlation, we also want to ask whether the monogamy relation can be used to investigate some fundamental physics problems, such as quantum phase transitions.

Quantum  phase transitions (QPT) \cite{gu,sachdev,chen2,gu2} indicates that the ground state of a many-body system changes abruptly when varying a physical parameter〞such as magnetic field or pressure at absolute zero temperature. Contrary to thermal phase transitions, QPT is completely induced by quantum fluctuations. Generally, researchers adopt order parameter, correlation functions and other concepts in thermal phase transitions to investigate QPT. Though many meaningful results have been got, there is still some shortage in it. The rapid development of quantum information science provides us many good means to understand the nature of QPT. A lot of studies \cite{osborne,osterloh,vidal1,huang,verstraete} indicate that the concept of entanglement and quantum correlation can be used to detect QPT or describe the property near the critical point. In addition, the renormalization group theory employed as one important measure to study QPT for many years. Recently, researchers begun to study QPT in low-dimensional spin systems by combing quantum information concepts and quantum renormalization group (QRG) theory. It has been shown that the behavior of the entanglement in the vicinity of the critical point is directly connected to the quantum critical properties \cite{kargarian1,kargarian2,jafari,ma1,ma2}. The quantum correlation measures also can be used to detect the quantum critical points associated with QPT after implementation of the QRG method \cite{yao,qin2,song}. However, these study mainly concentrate on one-dimensional systems. The two-dimensional spin systems, such as lanar quadrilateral spin crystal, triangular spin grid, and kagome spin lattice also are very important low-dimensional systems. The research on these models will improve the understanding of ground-state properties, correlation length, and critical point in two-dimensional system.

Recently, Xu \cite{xu} investigated the quantum entanglement around the quantum critical point for the Ising model on a square lattice. Usman \cite{usman} given an analysis of two-dimensional XY system by using entanglement theory. Nevertheless, as mentioned before, quantum entanglement is not adequate to represent all the quantum correlation contained in a quantum system, this inspires us to apply the quantum correlation measures to study such two-dimensional system \cite{yao}. Furthermore, question concerning monogamy relation also deserves our attention. Whether the monogamy  exists in two-dimensional spin system? Whether the monogamy relation can be used to catch the quantum critical point and give some useful tool to demonstrate it? To answer these questions, the dynamics behavior and the monogamy property of two-dimensional XY spin system is studied by the quantum correlation measures.
\section*{Model Hamiltonian}
The Hamiltonian of a two-dimensional XY spin model reads
\begin{equation}
\label{eq1}
H(J,\gamma)=\frac{J}{4}\sum_{i=1}^{N}\sum_{j=1}^{N}[(1+\gamma)(\sigma_{i,j}^{x}\sigma_{i+1,j}^{x}+\sigma_{i,j}^{x}\sigma_{i,j+1}^{x})+(1-\gamma)(\sigma_{i,j}^{y}\sigma_{i+1,j}^{y}+\sigma_{i,j}^{y}\sigma_{i,j+1}^{y})],
\end{equation}
where $J$ is the exchange interaction, $\gamma$ is the anisotropy parameter and $\sigma^{\tau} (\tau=x,y)$ are standard Pauli operators at site $i,j$. In order to apply QRG, we need to select five-site as one block. Such five-site blocks is viewed as one-site in renormalized subspace. The diagram is shown in Fig. 1. So, the Hamiltonian can be separated as block Hamiltonian $H^{B}$ and interacting Hamiltonian $H^{BB}$ respectively.
 \begin{equation}
\label{eq2}
\begin{split}
H^{B}=\frac{J}{4}\sum_{L=1}^{N/5}[(1+\gamma)(\sigma_{L,1}^{x}\sigma_{L,2}^{x}+\sigma_{L,1}^{x}\sigma_{L,3}^{x}+
\sigma_{L,1}^{x}\sigma_{L,4}^{x}+\sigma_{L,1}^{x}\sigma_{L,5}^{x})+\\
(1-\gamma)(\sigma_{L,1}^{y}\sigma_{L,2}^{y}+\sigma_{L,1}^{y}\sigma_{L,3}^{y}+\sigma_{L,1}^{y}\sigma_{L,4}^{y}+\sigma_{L,1}^{y}\sigma_{L,5}^{y})],
\end{split}
\end{equation}

\begin{equation}
\label{eq3}
\begin{split}
&H^{BB}=\frac{J}{4}\sum_{L=1}^{N/5}[(1+\gamma)(\sigma_{L,2}^{x}\sigma_{L+1,3}^{x}+\sigma_{L,2}^{x}\sigma_{L+1,4}^{x}+
\sigma_{L,2}^{x}\sigma_{L+2,5}^{x}+\sigma_{L,3}^{x}\sigma_{L+2,4}^{x}+\sigma_{L,3}^{x}\sigma_{L+2,5}^{x}+
\sigma_{L,4}^{x}\sigma_{L+3,5}^{x})  \\
&~~~~~~~~~~~~+(1-\gamma)(\sigma_{L,2}^{y}\sigma_{L+1,3}^{y}+\sigma_{L,2}^{y}\sigma_{L+1,4}^{y}+
\sigma_{L,2}^{y}\sigma_{L+2,5}^{y}+\sigma_{L,3}^{y}\sigma_{L+2,4}^{y}+\sigma_{L,3}^{y}\sigma_{L+2,5}^{y}+
\sigma_{L,4}^{y}\sigma_{L+3,5}^{y})].
\end{split}
\end{equation}

 The two lowest eigenvectors of the corresponding $L$-th block
 \begin{equation}
\label{eq4}
\begin{split}
|\varphi_{0}\rangle=\gamma_{1}(|\uparrow\uparrow\uparrow\uparrow\downarrow\rangle+|\uparrow\uparrow\uparrow\downarrow\uparrow\rangle+|\uparrow\uparrow\downarrow\uparrow\uparrow\rangle+
|\uparrow\downarrow\uparrow\uparrow\uparrow\rangle)+ \\
\gamma_{2}(|\uparrow\uparrow\downarrow\downarrow\downarrow\rangle+|\uparrow\downarrow\uparrow\downarrow\downarrow\rangle+
|\uparrow\downarrow\downarrow\uparrow\downarrow\rangle+|\uparrow\downarrow\downarrow\downarrow\uparrow\rangle)+\\
\gamma_{3}|\downarrow\uparrow\uparrow\uparrow\uparrow\rangle+\gamma_{4}(|\downarrow\uparrow\uparrow\uparrow\uparrow\rangle+|\downarrow\uparrow\uparrow\downarrow\downarrow\rangle+
|\downarrow\uparrow\downarrow\downarrow\uparrow\rangle+\\
|\downarrow\downarrow\uparrow\uparrow\downarrow\rangle+|\downarrow\downarrow\uparrow\downarrow\uparrow\rangle+
|\downarrow\downarrow\downarrow\uparrow\uparrow\rangle)+\gamma_{5}|\downarrow\downarrow\downarrow\downarrow\downarrow\rangle),
\end{split}
\end{equation}
and
 \begin{equation}
\label{eq5}
\begin{split}
|\varphi_{0}^{'}\rangle=\gamma_{6}|\uparrow\uparrow\uparrow\uparrow\uparrow\rangle+\gamma_{7}(|\uparrow\uparrow\uparrow\downarrow\downarrow\rangle+|\uparrow\uparrow\downarrow\uparrow\downarrow\rangle+|\uparrow\uparrow\downarrow\downarrow\uparrow\rangle\\
+|\uparrow\downarrow\uparrow\uparrow\downarrow\rangle+|\uparrow\downarrow\uparrow\downarrow\uparrow\rangle+|\uparrow\downarrow\downarrow\uparrow\uparrow\rangle)+\gamma_{8}|\uparrow\downarrow\downarrow\downarrow\downarrow\rangle\\
+\gamma_{9}(|\downarrow\uparrow\uparrow\uparrow\downarrow\rangle+|\downarrow\uparrow\uparrow\downarrow\uparrow\rangle+|\downarrow\uparrow\downarrow\uparrow\uparrow\rangle+|\downarrow\downarrow\uparrow\uparrow\uparrow\rangle)\\
+\gamma_{10}(|\downarrow\uparrow\downarrow\downarrow\downarrow\rangle+|\downarrow\downarrow\uparrow\downarrow\downarrow\rangle+|\downarrow\downarrow\downarrow\uparrow\downarrow\rangle+|\downarrow\downarrow\downarrow\downarrow\uparrow\rangle),
\end{split}
\end{equation}
 can be used to establish the projection operator $P_{0}^{L}=|\varphi_{0}\rangle_{L}\langle\Uparrow|+|\varphi_{0}^{'}\rangle_{L}\langle\Downarrow|$. Analytical expressions of the $\gamma_{i}$＊s are in Ref. \cite{usman}, $\langle\Uparrow|, \langle\Downarrow|$ are renamed states of each block to represent the effective site degrees of freedom. The effective Hamiltonian is given by
\begin{equation}
\label{eq6}
\begin{split}
& H^{eff}(J^{'},\gamma^{'})=P_{0}^{\dag}(H^{B}+H^{BB})P_{0}\\
&~~~~~~~~~~~~~~~=\frac{J^{'}}{4}\sum_{p=1}^{N/5}\sum_{q=1}^{N/5}[(1+\gamma^{'})(\sigma_{p,q}^{x}\sigma_{p+1,q}^{x}+\sigma_{p,q}^{x}\sigma_{p,q+1}^{x})+(1-\gamma^{'})(\sigma_{p,q}^{y}\sigma_{p+1,q}^{y}+\sigma_{p,q}^{y}\sigma_{p,q+1}^{y})],
\end{split}
\end{equation}
where the renormalized couplings are

\begin{equation}
\label{eq7}
\begin{split}
&J^{'} = J(\gamma_{10}^{2}(9\gamma_{4}^{2}+6\gamma\gamma_{4}\gamma_{5}+\gamma_{5}^{2})+9\gamma_{2}^{2}\gamma_{7}^{2}
+\gamma_{1}^{2}(\gamma_{6}^{2}+6\gamma\gamma_{6}\gamma_{7}+9\gamma_{7}^{2})+6\gamma\gamma_{2}^{2}\gamma_{7}\gamma_{8}
+\gamma_{2}^{2}\gamma_{8}^{2}+6\gamma\gamma_{2}\gamma_{3}\gamma_{7}\gamma_{9}\\
&~~~~+18\gamma\gamma_{2}\gamma_{4}\gamma_{7}\gamma_{9}+2\gamma_{2}\gamma_{3}\gamma_{8}\gamma_{9}+6\gamma\gamma_{2}\gamma_{4}\gamma_{8}\gamma_{9}
+\gamma_{3}^{2}\gamma_{9}^{2}+6\gamma\gamma_{3}\gamma_{4}\gamma_{9}^{2}+9\gamma_{4}^{2}\gamma_{9}^{2}
+2\gamma_{1}\{\gamma_{2}[3\gamma_{7}(3\gamma\gamma_{7}+\gamma_{8})\\
&~~~~+\gamma_{6}(3\gamma_{7}+\gamma\gamma_{8})]+(\gamma\gamma_{3}\gamma_{6}+3\gamma_{4}\gamma_{6}+3\gamma_{3}\gamma_{7}
+9\gamma\gamma_{4}\gamma_{7})\gamma_{9}\}
+2\gamma_{10}\{\gamma_{1}(\gamma_{5}\gamma_{6}+9\gamma_{4}\gamma_{7})+\gamma[9\gamma_{2}\gamma_{4}\gamma_{7}\\
&~~~~+3\gamma_{1}(\gamma_{4}\gamma_{6}+\gamma_{5}\gamma_{7})+\gamma_{2}\gamma_{5}\gamma_{8}
+9\gamma_{4}^{2}\gamma_{9}+\gamma_{3}\gamma_{5}\gamma_{9}]+3[\gamma_{2}(\gamma_{5}\gamma_{7}
+\gamma_{4}\gamma_{8})+\gamma_{4}(\gamma_{3}+\gamma_{5})\gamma_{9}]\}),
\end{split}
\end{equation}
and
\begin{equation}
\label{eq8}
\begin{split}
&\gamma^{'}=[2(3\gamma_{10}\gamma_{4}+3\gamma_{1}\gamma_{7}+\gamma_{2}\gamma_{8}+\gamma_{3}\gamma_{9})(\gamma_{10}\gamma_{5}
+\gamma_{1}\gamma_{6}+3\gamma_{2}\gamma_{7}+3\gamma_{4}\gamma_{9})+\gamma(\gamma_{10}^{2}(9\gamma_{4}^{2}+\gamma_{5}^{2})
+9\gamma_{2}^{2}\gamma_{7}^{2}\\
&~~~~+\gamma_{1}^{2}(\gamma_{6}^{2}+9\gamma_{7}^{2})+\gamma_{2}^{2}\gamma_{8}^{2}
+18\gamma_{2}\gamma_{4}\gamma_{7}\gamma_{9}+2\gamma_{2}\gamma_{3}\gamma_{8}\gamma_{9}+\gamma_{3}^{2}\gamma_{9}^{2}
+6\gamma_{1}[\gamma_{2}\gamma_{7}(\gamma_{6}+\gamma_{8})+(\gamma_{4}\gamma_{6}+\gamma_{3}\gamma_{7})\gamma_{9}]\\
&~~~~+2\gamma_{10}\{\gamma_{1}(\gamma_{5}\gamma_{6}+9\gamma_{4}\gamma_{7})+3[\gamma_{2}(\gamma_{5}\gamma_{7}+\gamma_{4}\gamma_{8})
+\gamma_{4}(\gamma_{3}+\gamma_{5})\gamma_{9}]\})]/[\gamma_{10}^{2}(9\gamma_{4}^{2}+6\gamma\gamma_{4}\gamma_{5}+\gamma_{5}^{2})
+9\gamma_{2}^{2}\gamma_{7}^{2}\\
&~~~~+\gamma_{1}^{2}(\gamma_{6}^{2}+6\gamma\gamma_{6}\gamma_{7}+9\gamma_{7}^{2})
+6\gamma\gamma_{2}^{2}\gamma_{7}\gamma_{8}+\gamma_{2}^{2}\gamma_{8}^{2}+6\gamma\gamma_{2}\gamma_{3}\gamma_{7}\gamma_{9}
+18\gamma_{2}\gamma_{4}\gamma_{7}\gamma_{9}+2\gamma_{2}\gamma_{3}\gamma_{8}\gamma_{9}+6\gamma\gamma_{2}\gamma_{4}\gamma_{8}\gamma_{9}\\
&~~~~+\gamma_{3}^{2}\gamma_{9}^{2}+6\gamma\gamma_{3}\gamma_{4}\gamma_{9}^{2}+9\gamma_{4}^{2}\gamma_{9}^{2}+2\gamma_{1}\{\gamma_{2}[
3\gamma_{7}(3\gamma\gamma_{7}+\gamma_{8})+\gamma_{6}(3\gamma_{7}+\gamma\gamma_{8})]+(\gamma\gamma_{3}\gamma_{6}+3\gamma_{4}\gamma_{6}\\
&~~~~+3\gamma_{3}\gamma_{7}+9\gamma\gamma_{4}\gamma_{7})\gamma_{9}\}+2\gamma_{10}(\gamma_{1}(\gamma_{5}\gamma_{6}+9\gamma_{4}\gamma_{7})
+\gamma[9\gamma_{2}\gamma_{4}\gamma_{7}+3\gamma_{1}(\gamma_{4}\gamma_{6}+\gamma_{5}\gamma_{7})+\gamma_{2}\gamma_{5}\gamma_{8}\\
&~~~~+9\gamma_{4}^{2}\gamma_{9}+\gamma_{3}\gamma_{5}\gamma_{9}]+3[\gamma_{2}(\gamma_{5}\gamma_{7}+\gamma_{4}\gamma_{8})
+\gamma_{4}(\gamma_{3}+\gamma_{5})\gamma_{9}])].
\end{split}
\end{equation}
The ground state density matrix is
\begin{equation}
\label{eq9}
\rho=|\varphi_{0}\rangle\langle\varphi_{0}|.
\end{equation}
Owing to the symmetry, the bipartite state between the center block and every corners block is identical that is $\rho_{12}=\rho_{13}=\rho_{14}=\rho_{15}$. Similarly, $\rho_{23}=\rho_{34}=\rho_{45}=\rho_{25}$.
After some algebra, we derive the bipartite state $\rho_{12}$ and $\rho_{23}$ by tracing out particles 3,4,5 or 1,4,5.

\begin{equation}
\label{eq10}
\rho_{12}=\left(
               \begin{array}{cccc}
                 3\gamma_{1}^{2}+\gamma_{2}^{2} & 0 & 0 & 3\gamma_{1}\gamma_{4}+\gamma_{2}\gamma_{5} \\
                 0 & \gamma_{1}^{2}+3\gamma_{2}^{2} & \gamma_{1}\gamma_{3}+3\gamma_{2}\gamma_{4} & 0 \\
                 0 & \gamma_{1}\gamma_{3}+3\gamma_{2}\gamma_{4} & \gamma_{3}^{2}+3\gamma_{4}^{2} & 0 \\
                 3\gamma_{1}\gamma_{4}+\gamma_{2}\gamma_{5} & 0 & 0 & 3\gamma_{4}^{2}+\gamma_{5}^{2} \\
               \end{array}
             \right),
\end{equation}

\begin{equation}
\label{eq11}
\rho_{23}=\left(
               \begin{array}{cccc}
                 2\gamma_{1}^{2}+\gamma_{3}^{2}+\gamma_{4}^{2} & 0 & 0 & 2\gamma_{1}\gamma_{2}+\gamma_{3}\gamma_{4}+\gamma_{4}\gamma_{5} \\
                 0 & \gamma_{1}^{2}+\gamma_{2}^{2}+2\gamma_{4}^{2} & \gamma_{1}^{2}+\gamma_{2}^{2}+2\gamma_{4}^{2} & 0 \\
                 0 & \gamma_{1}^{2}+\gamma_{2}^{2}+2\gamma_{4}^{2} & \gamma_{1}^{2}+\gamma_{2}^{2}+2\gamma_{4}^{2} & 0 \\
                 2\gamma_{1}\gamma_{2}+\gamma_{3}\gamma_{4}+\gamma_{4}\gamma_{5} & 0 & 0 & 2\gamma_{2}^{2}+\gamma_{4}^{2}+\gamma_{5}^{2} \\
               \end{array}
             \right).
\end{equation}

\section*{Results}
Next, we use quantum correlation measures and quantum nonlocality measure as well as Bell violation to investigate the quantum critical properties of this model.

\textbf{Negativity.}
Negativity ($Ne$) \cite{vidal2} is an easily computable entanglement measure, it was introduced for testing the violation degree of positive partial transpose criterion in entangled states. $Ne$ was proved to be monotone and convex under local operations and classical communication. For a bipartite system $\rho_{AB}$, the partial transpose of $\rho_{AB}$ on $A$ can be described as $(\rho_{AB}^{T_{A}})_{ij,kl}=(\rho)_{kj,il}$. So, for a given state $\rho_{AB}$, the $Ne$ is
\begin{equation}
\label{eq12}
Ne(\rho_{AB})=\frac{\|\rho_{AB}^{T_{A}}\|-1}{2},
\end{equation}
where $\|\rho_{AB}^{T_{A}}\|=Tr\sqrt{\rho_{AB}^{T_{A}}\rho_{AB}^{T_{A}\dagger}}$ denotes the trace norm. For the bipartite $2\otimes2$ and $2\otimes3$ quantum systems, $Ne$ is the necessary and sufficient inseparable condition.

The analytical results of $Ne$ of states $\rho_{12}$ and $\rho_{13}$ are
\begin{equation}
\label{eq13}
\begin{split}
&Ne_{12}=(9\gamma_{1}^{4}+6\gamma_{1}^{2}\gamma_{2}^{2}+4\gamma_{1}^{2}\gamma_{3}^{2}-18\gamma_{1}^{2}\gamma_{4}^{2}-6\gamma_{1}^{2}\gamma_{5}^{2}
+24\gamma_{1}\gamma_{2}\gamma_{3}\gamma_{4}+\gamma_{2}^{4}+30\gamma_{2}^{2}\gamma_{4}^{2}-2\gamma_{2}^{2}\gamma_{5}^{2}+9\gamma_{4}^{4}\\
&+6\gamma_{4}^{2}\gamma_{5}^{2}+\gamma_{5}^{4})^{1/2}/ 2
-(\gamma_{2}^{2}+3\gamma_{4}^{2}+\gamma_{5}^{2}+3\gamma_{1}^{2})/2,
\end{split}
\end{equation}
and
\begin{equation}
\label{eq14}
\begin{split}
&Ne_{23}=(8\gamma_{1}^{4}+4\gamma_{1}^{2}\gamma_{3}^{2}+16\gamma_{1}^{2}\gamma_{4}^{2}-4\gamma_{1}^{2}\gamma_{5}^{2}+8\gamma_{2}^{4}
-4\gamma_{2}^{2}\gamma_{3}^{2}+16\gamma_{2}^{2}\gamma_{4}^{2}+4\gamma_{2}^{2}\gamma_{5}^{2}+\gamma_{3}^{4}-2\gamma_{3}^{2}\gamma_{5}^{2}\\
&+16\gamma_{4}^{4}+\gamma_{5}^{4})^{1/2}/2-(\gamma_{1}^{2}+\gamma_{2}^{2}+\gamma_{3}^{2}/2+\gamma_{4}^{2}+\gamma_{5}^{2}/2).
\end{split}
\end{equation}
\textbf{Quantum discord.}
As we know, the measure $Ne$ is based on entanglement-separablity paradigm. But Quantum discord ($QD$) is proposed from the perspective of information-theoretic paradigm. The expression takes the form \cite {ollivier}
\begin{equation}
\label{eq15}
QD(\rho_{AB})=I(\rho_{AB})-CC(\rho_{AB}),
\end{equation}
where $I(\rho_{AB})$ is the total correlation and measured by the quantum mutual information $I(\rho_{AB})=\sum_{i=A,B}S(\rho_{i})-S(\rho_{AB})$, while $CC(\rho_{AB})=S(\rho_{A})-\rm min_{\Pi_{k}^{B}} S(S_{A|B}\{\Pi_{k}^{B}\})$, in which $\Pi_{k}^{B}$ is a positive operator-valued measue performed on the subsystem $B$. $S(\rho)=-Tr(\rho\log_{2}\rho)$ is the von Neumann entropy, $\rho_{A}$ and $\rho_{B}$ denote the reduced density matrix of state $\rho_{AB}$ by tracing out $A$ or $B$.

Since $\rho_{12}$ and $\rho_{13}$  are X-type states, it is easy to compute the quantum discord result \cite{chen,yu,ali,huang2}. Even so, the analytical result of this case is too complicated to express it here. We mainly show the numerical result of QD in section IV.

\textbf{Measurement-induced disturbance}
Measurement-induced disturbance ($MID$) \cite{luo} was defined by the difference between the two quantum mutual information of a quantum state $\rho_{AB}$ and the corresponding post-measurement classical state $\Pi(\rho_{AB})$
\begin{equation}
\label{eq16}
MID(\rho_{AB})=I(\rho_{AB})-I(\Pi(\rho_{AB})),
\end{equation}
here $I(\rho_{AB})$ is the same as in Eq. (15). $I(\Pi(\rho_{AB}))=\sum_{i,j}(\Pi_{i}^{A}\otimes\Pi_{i}^{B})\rho_{AB}(\Pi_{i}^{A}\otimes\Pi_{i}^{B})$ measures the classical correlation in a given state $\rho_{AB}$.

The analytical result of the $MID$ 0f $\rho_{12}$ and $\rho_{13}$ are also very complicated and we shall not write it here. People can derive the eigenvalues and the diagonal element of $\rho_{12}$ and $\rho_{13}$, and then deduce the results of $MID$.

\textbf{Measurement-induced nonlocality}
The measurement-induce nonlocality ($MIN$) bases on the trace norm for a bipartite state $\rho_{AB}$ expressed as \cite{hu}
\begin{equation}
\label{eq17}
MIN(\rho_{AB})=\rm max_{\Pi^{A}}\|\rho_{AB}-\Pi^{A}(\rho_{AB})\|_{1},
\end{equation}
here $\|R\|_{1}=Tr\sqrt{R^{\dag}R}$, and the maximum is taken over the full set of local projective measurements $\Pi^{A}$  that $\Pi^{A}(\rho_{A})=\rho_{A}$.

The analytical result of $MIN$ of $\rho_{12}$ and $\rho_{13}$ given by
\begin{equation}
\label{eq18}
MIN_{12}=\rm max(|t_{1}|,|t_{2}|,|t_{3}|),
\end{equation}
\begin{equation}
\label{eq19}
MIN_{23}=(|l_{1}-l_{2}|+|l_{1}+l_{2}|)/2,
\end{equation}
where $t_{1}=2\gamma_{1}\gamma_{3}+6\gamma_{4}(\gamma_{1}+\gamma_{2})+2\gamma_{2}\gamma_{5}$,
$t_{2}=2\gamma_{1}\gamma_{3}+6\gamma_{4}(\gamma_{2}-\gamma_{1})-2\gamma_{2}\gamma_{5}$,
$t_{3}=2\gamma_{1}^{2}-2\gamma_{2}^{2}-\gamma_{3}^{2}+\gamma_{5}^{2}$,
$l_{1}=2(\gamma_{1}+\gamma_{2})^{2}+2\gamma_{4}(\gamma_{3}+\gamma_{5})+4\gamma_{4}^{2}$,
$l_{2}=2(\gamma_{1}-\gamma_{2})^{2}-2\gamma_{4}(\gamma_{3}+\gamma_{5})+4\gamma_{4}^{2}$,
$l_{3}=\gamma_{3}^{2}-2\gamma_{4}^{2}+\gamma_{5}^{2}$.

\textbf{Geometric quantum discord}
The geometric measure of quantum discord ($GQD$) is taken to be \cite{dakic}
\begin{equation}
\label{eq20}
GQD(\rho_{AB}):=\rm min_{x\in\Omega}\|\rho-\chi\|,
\end{equation}
where $\Omega$ means the set of zero-discord states, whose general form is defined by $\chi=\sum_{k}p_{k}\Pi_{k}^{A}\otimes\rho_{k}^{B}$ with $0\leq p_{k} \leq 1 (\sum_{k}p_{k}=1)$, and $\|\rho-\chi\|^{2}=tr(\rho-\chi)^{2}$ means the square of the Hilbert-Schmidt
norm.

The analytical results of the present model reads
\begin{equation}
\label{eq21}
GQD_{12}=(t_{1}^{2}+t_{2}^{2}+t_{3}^{2}+x_{1}^{2}-max(t_{1}^{2},t_{2}^{2},t_{3}^{2}+x_{1}^{2}))/4,
\end{equation}
\begin{equation}
\label{eq22}
GQD_{23}=(l_{1}^{2}+l_{2}^{2}+l_{3}^{2}+x_{2}^{2}-max(l_{1}^{2},l_{2}^{2},l_{3}^{2}+x_{2}^{2}))/4,
\end{equation}
where $t_{1},t_{2},t_{3},l_{1},l_{2}, l_{3}$ are the same as Eq. (18) and Eq. (19), and $x_{1}=4\gamma_{1}^{2}+4\gamma_{2}^{2}-\gamma_{3}^{2}-6\gamma_{4}^{2}-\gamma_{5}^{2}$,
$x_{2}=2\gamma_{1}^{2}-2\gamma_{2}^{2}+\gamma_{3}^{2}-\gamma_{5}^{2}$.

\textbf{Bell violation}
The violation of Bell inequality is accepted as the existence of quantum nonlocality. Following equation is the Bell operator corresponding to the Clauser-Horne-Shimony-Holt (CHSH) form \cite{horodecki}
\begin{equation}
\label{eq23}
B=a\cdot\sigma\otimes(b+b^{'})\cdot\sigma+a^{'}\cdot\sigma\otimes(b-b^{'})\cdot\sigma
\end{equation}
where $a, a^{'}, b, b^{'}$ are the unit vectors in $\mathbb{R}^{3}$, and the CHSH inequality can be written as $B=|\langle B_{CHSH}\rangle|=|Tr(\rho_{CHSH})|\leq 2$, in which the maximum violation of CHSH inequality obey
\begin{equation}
\label{eq24}
B_{CHSH}^{max}=max_{a,a',b,b'}Tr(\rho B_{CHSH}).
\end{equation}

The analytical result of $B$ for $\rho_{12}$ and $\rho_{13}$ are given by
\begin{equation}
\label{eq25}
B_{12}=\sqrt{t_{1}^{2}+t_{2}^{2}+t_{3}^{2}-\lambda_{min}},
\end{equation}
\begin{equation}
\label{eq26}
B_{23}=\sqrt{l_{1}^{2}+l_{2}^{2}+l_{3}^{2}-\kappa_{min}},
\end{equation}
here $t_{1},t_{2},t_{3},l_{1},l_{2}, l_{3}$ also are the same as before, and $\lambda_{min}=min(t_{1}^{2},t_{2}^{2},t_{3}^{2})$,
$\kappa_{min}=min(l_{1}^{2},l_{2}^{2},l_{3}^{2})$.

\textbf{Dynamics behavior of different quantum correlation.}
According to above-mentioned quantum correlation measures, the dynamics behavior of every quantity can be gotten by implementing QRG method.

The properties of different quantum correlation measures versus  $\gamma$  in terms of QRG iterations are plotted in Fig. 2. After two steps of renormalization, $Ne$ will develop two saturated values, one that is nonzero for $\gamma_{c}$=0 and one that is zero for $\gamma_{c}\neq$ 0. $QD$ and $GQD$ have the same property as entanglement. But the fixed value of $MID$ and $MIN$ are a little different, namely one that is nonzero for $\gamma_{c}$=0 and one that is 1 for $\gamma_{c}\neq$ 0, the Bell inequality also have the same characteristic. The plots cross each other at the critical point $\gamma_{c}$=0, which means that the block-block correlations of $\rho_{12}$ will demonstrate QPT at the critical point $\gamma_{c}$=0. Moreover, this state cannot violate the CHSH inequality.

In Fig. 3, we illustrate the quantum correlation evolution of $\rho_{23}$  versus $\gamma$ for different QRG steps. The quantum correlation in $\rho_{23}$ is  between the two corner-site blocks.  The behavior of every quantum correlation measures are roughly the same with Fig. 2 but also have small difference, such as the saturated value and the change rate vs $\gamma$.

\textbf{Nonanalytic and scaling behavior.}
The first derivative of different quantum correlation measures (DQCM) vs. $\gamma$ are shown in Fig. 4. From
this figure, we notice that the derivative of quantum correlation diverges at the critical point $\gamma$ = 0 \cite{jafari}. All the plots in the figure exhibit as an antisymmetrical function about $\gamma$ = 0 . There is a maximum and a minimum value for each plot, and the peak value becomes more pronounced near to the critical point $\gamma$ = 0. This indicates that the two-dimensional XY system displays a second-order QPT. Comparing the six subgraphs, we note that the absolute peak value of $QD$ and $MIN$ are larger than the other measures. This implies that the $QD$ and $MIN$ are more sensitive than the other quantum correlation measures to detect QPT.

Figure 5 shows the first derivatives of the quantum correlation measures as a function of $\gamma$  after tracing out block 1, 4, and 5.  From the figure, it is immediately seen that the change rate and peak value of $\rho_{23}$ is substantially quicker and larger than Fig. 4. Specifically, when $MID$ and $MIN$ is adopted, the absolute value of the first derivative of $\rho_{23}$ are reaching 200. This singular behavior represents a more sensitive and more pronounced property close to the critical point $\gamma$ = 0 for state $\rho_{23}$.

We demonstrate that the first derivative of different quantum correlation measures show the nonanalytic behavior at the critical point. A more detailed analysis shows that the maximum and minimum of the first derivative exhibit the scaling behavior versus $N$. Results are presented in Fig. 6, which displays a linear behavior of $\ln dDQCM/d\gamma$ versus $\ln N$. The scaling behavior is approximately $dDQCM/d\gamma_{max} \sim N^{1.13}$ or $dDQCM/d\gamma_{min}\sim N^{1.13}$. One of the important results that we conclude from our figure is that the exponent $\theta$ are generally identical, which means that the exponent will not change with the variation of quantum correlation measures. Since the critical exponent directly associates with the correlation length exponent \cite{jafari}, this result establish the relation between quantum information theory and condensed matter physics.

\textbf{The monogamy relation of two dimensional XY modle.}
Monogamy relation of entanglement \cite{coffman} has been a subject in the quantum information processing over the years. It is worthwhile to investigate whether the monogamy property of entanglement and quantum correlation exist in this two-dimensional system? Another question is whether the monogamy relation can be used to detect QPT? Here we will select two typical measures that is concurrence and $QD$ as the quantity. The concurrence ($C$) \cite{hill} of a bipartite state is $C=Max\{\lambda_{1}-\lambda_{2}-\lambda_{3}-\lambda_{4},0\}$, where $\lambda_{k}( k=1,2,3,4)$  are the square roots of the eigenvalues in descending order of the operator $R_{AB}=\rho_{AB}\widetilde{\rho}_{AB}$, $\widetilde{\rho}_{AB}=(\sigma_{1}^{y}\sigma_{2}^{y})\rho_{AB}^{\ast}(\sigma_{1}^{y}\sigma_{2}^{y})$.  For this five-site block state, two kinds of the inequality in terms of concurrence are expressed by \cite{ou,song2,luo2,cornelio}
\begin{equation}
\label{eq27}
C_{12}^{2}+C_{13}^{2}+C_{14}^{2}+C_{15}^{2}\leq C_{1|2345}^{2},
\end{equation}
\begin{equation}
\label{eq28}
C_{21}^{2}+C_{23}^{2}+C_{24}^{2}+C_{25}^{2}\leq C_{2|1345}^{2},
\end{equation}
where $C_{ij}$ stands for the concurrence of the density matrix $\rho$ with blocks other than $i,j$ traced out, and $C_{i|jklm}$ stands for the concurrence between the subsystems $\rho_{i}$ and $\rho_{jklm}$. The analytical expressions of $C_{ij}$ can be computed through the above formula and $C_{i|jklm}=2\sqrt{\rho_{i}}$. The difference between the two sides of inequality set as the residual entanglement that is $\delta_{1(2345)}=C_{1|2345}^{2}-C_{12}^{2}-C_{13}^{2}-C_{14}^{2}-C_{15}^{2}$ and $ \delta_{2(1345)}=C_{2|1345}^{2}-C_{21}^{2}-C_{23}^{2}-C_{24}^{2}-C_{25}^{2}$.

Similarly, we can derive the monogamy inequality of $QD$ \cite{bai}
\begin{equation}
\label{eq29}
QD_{12}^{2}+QD_{13}^{2}+QD_{14}^{2}+QD_{15}^{2}\leq QD_{1|2345}^{2},
\end{equation}
\begin{equation}
\label{eq30}
QD_{21}^{2}+QD_{23}^{2}+QD_{24}^{2}+QD_{25}^{2}\leq QD_{2|1345}^{2},
\end{equation}
here $QD_{ij}$ have the similar meaning like concurrence but stand for the quantum correlation, and $QD_{i|jklm}=S(\rho_{i})$.
We also can define the difference between the two sides of inequality relation that is $\Delta_{1(2345)}=QD_{1|2345}^{2}-QD_{12}^{2}-QD_{13}^{2}-QD_{14}^{2}-QD_{15}^{2}$ and $ \Delta_{2(1345)}=QD_{2|1345}^{2}-QD_{21}^{2}-QD_{23}^{2}-QD_{24}^{2}-QD_{25}^{2}$.

Numerical simulations are performed for concurrence and $QD$ in Fig. 7, and we show that the concurrence and $QD$ are monogamous in two-dimensional XY system. The curves of $\delta_{1(2345)}$, $\delta_{2(1345)}$, $\Delta_{1(2345)}$, and $\Delta_{2(1345)}$  also cross each other at $\gamma_{c}=0$. This means that the residual entanglement and residual quantum correlation can be used to indicate the QPT. The difference of the monogamy relation which is defined as monogamy score can be regarded as a good tool to investigate the QPT. Furthermore, such monogamy score can characterize the genuine quantum correlation in this model\cite{bai}.

\section*{Discussion}
In this work we have studied the renormalization of entanglement, quantum correlation, and monogamy relation of two-dimensional XY model. As opposed to the one-dimensional case, the two-dimensional system size increases rapidly because we select five-site as one block. Furthermore, the critical point and the saturated values can be reached in the lesser number of QRG iterations. The scaling behavior have investigated through determination of the quantum correlation exponent which demonstrates how the critical point is attained as the size of the model becomes large. Remarkably, we have obtained the identical critical exponent of entanglement, quantum correlation and Bell-equality. Moreover, we have studied multipartite quantum correlations with the monogamy of concurrence and monogamy of quantum discord and shown that the two quantities are monogamous in this model. This studies will help us deeply understand the quantum critical problem in condensed matter physics.

\begin{acknowledgments}
This work was supported by the National Natural Science Foundation of China (Grant Nos 11535004, 11375086, 1175085, 11120101005 and 11235001), by the 973 National Major State Basic Research and Development of China grant No. 2013CB834400, and Technology Development Fund of Macau grant No. 068/2011/A., by the Project Funded by the Priority Academic Program Development of Jiangsu Higher Education Institutions (PAPD), by the Research and Innovation Project for College Postgraduate of JiangSu Province (Grants No. KYZZ15\_0027).
\end{acknowledgments}

\section*{Author contributions}
M.Q. carried out the calculations and wrote the paper. M.Q., Z.Z.R. and X.Z. discussed the results. All authors reviewed the manuscript.

\textbf{Competing financial interests:} The authors declare no competing financial interests.

\begin{figure}[H]
\centering
\includegraphics[width=9cm]{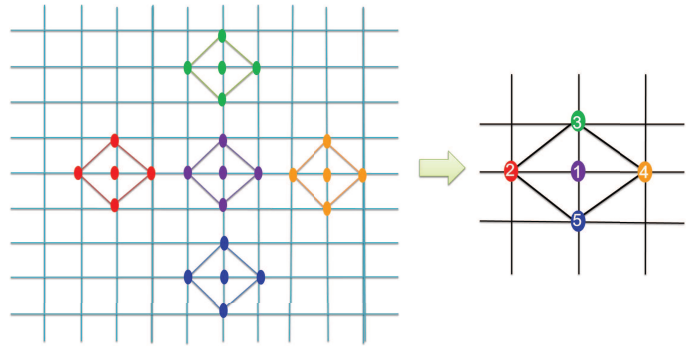}
\caption{A schematic description of QRG for five sites in a block.}\label{fig1}
\end{figure}

\begin{figure}[H]
\centering
\includegraphics[width=9cm]{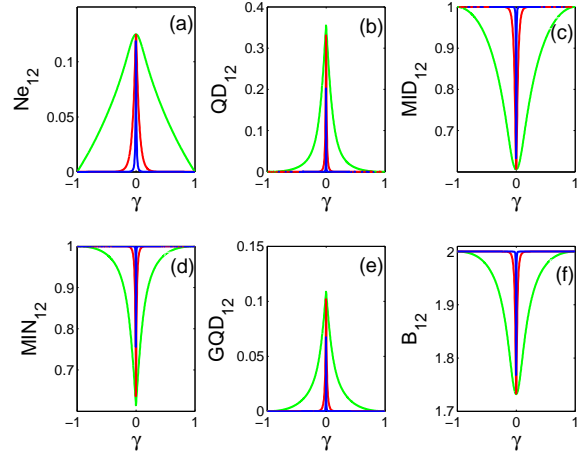}
\caption{Quantum correlation measures as a function of $\gamma$ at different QRG steps for state $\rho_{12}$. The green lines indicate 0-th step QRG, the red lines indicate 1-th step QRG, and the blue ones indicate 2-th step QRG.}
\end{figure}

\begin{figure}[H]
\centering
\includegraphics[width=9cm]{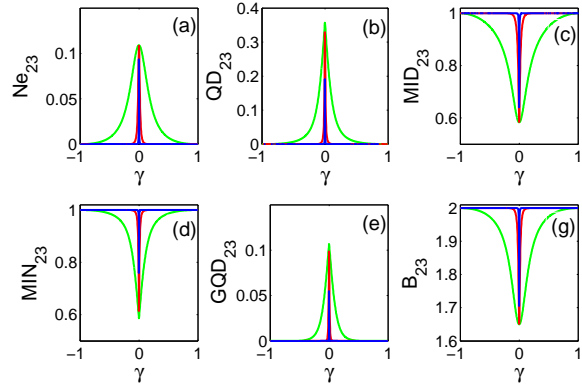}
\caption{Quantum correlation measures as a function of $\gamma$ at different QRG steps for state $\rho_{23}$. The plots color meaning are like before.}
\end{figure}

\begin{figure}[H]
\centering
\includegraphics[width=10cm]{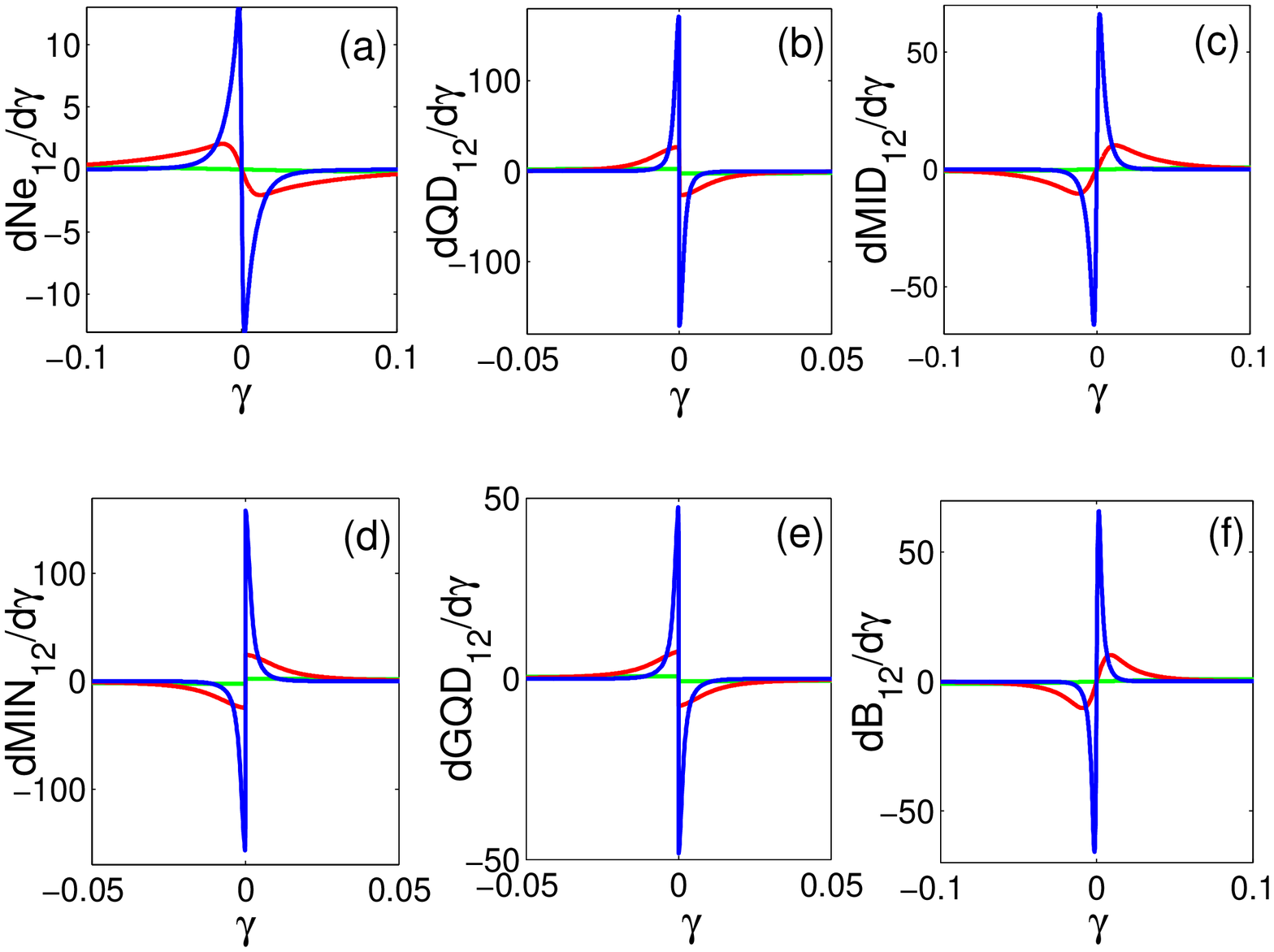}
\caption{Evolution of the first derivative of quantum correlation measures under QRG for state $\rho_{12}$. The plots color meaning are like before. }
\end{figure}

\begin{figure}[H]
\centering
\includegraphics[width=10cm]{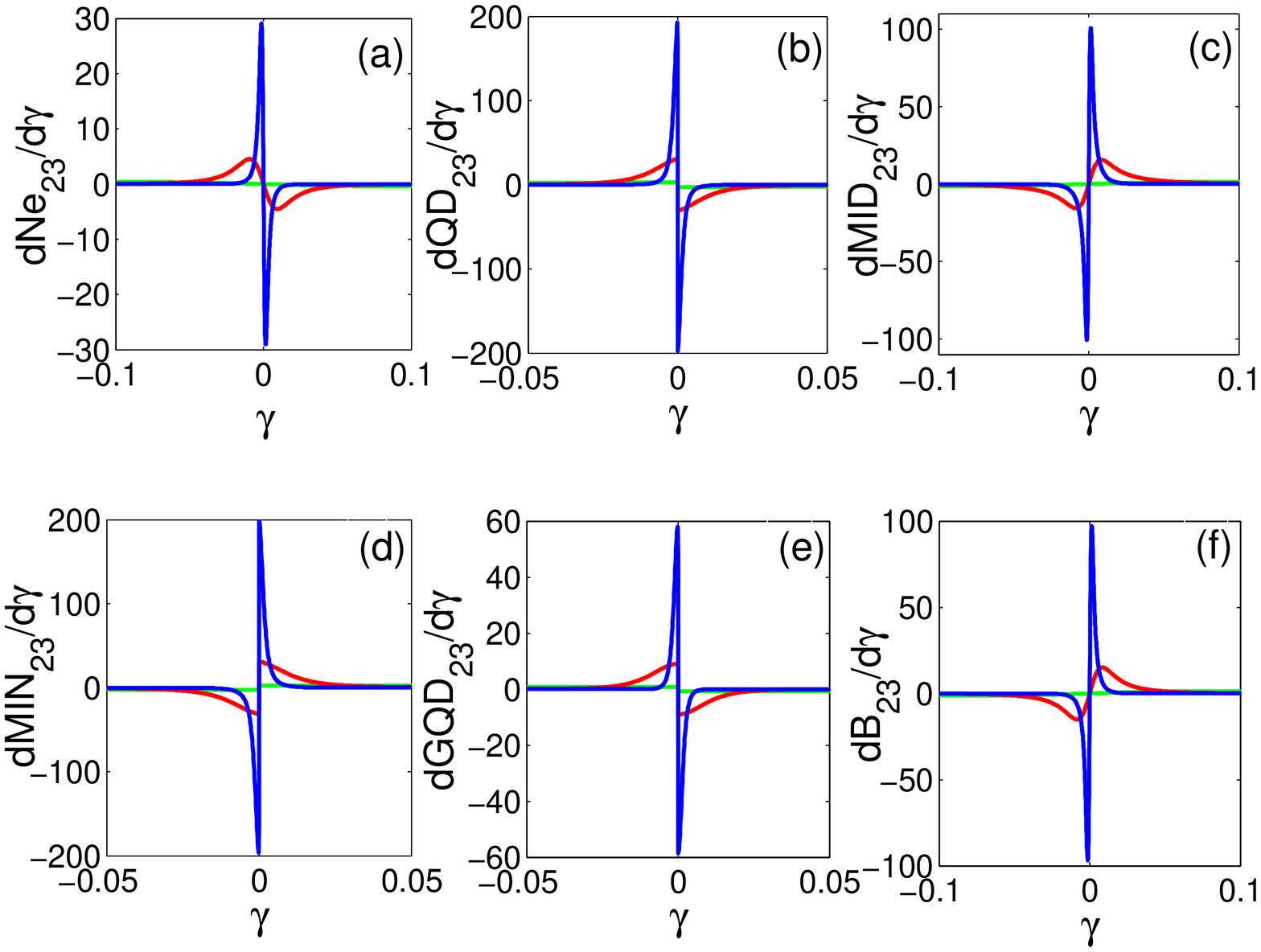}
\caption{Evolution of the first derivative quantum correlation measures under QRG for state $\rho_{23}$. The plots color meaning are like before. }
\end{figure}

\begin{figure}[H]
\centering
\includegraphics[scale=0.45]{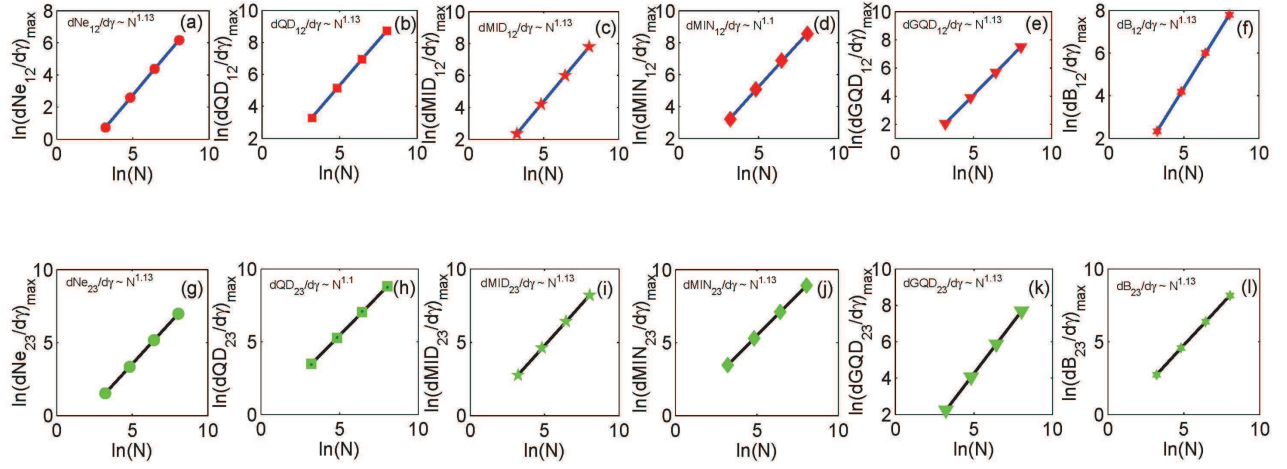}
\caption{The scaling behavior of different quantum correlations in terms of system size $\ln(N)$. }
\end{figure}

\begin{figure}[H]
\centering
\includegraphics[width=9cm]{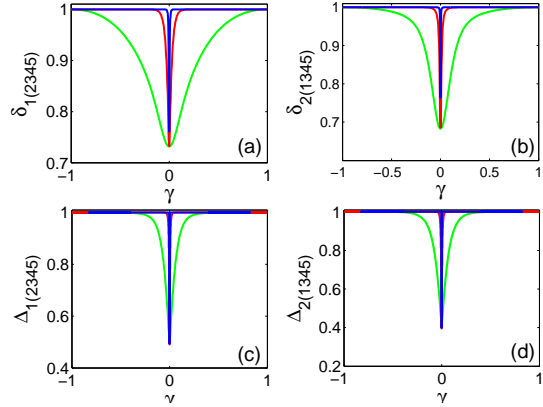}
\caption{The change of $\delta$ (a) and $\Delta$ (b) of the model versus $\gamma$ at different QRG
steps.  The plots color meaning are like before. }
\end{figure}

\end{document}